\documentclass[preprint2,longabstract,natbib,8pt]{aastex}

\usepackage{url}
\usepackage{here}
\usepackage{color}
\def\ADDED{}

\def\vect{\mathbf}
\def\pos{\vect r}
\def\uel{\vect u}   % Fluid Mean Velocity
\def\vel{\vect v}   % Particle Velocity, or Velocity Coordinate of Boltzmann Space
\def\uel{\vect u}
\def\mom{\vect p}

\def\Ele{\vect E}
\def\Mag{\vect B}

\def\max{\mathrm {max}}
\def\min{\mathrm {min}}
\def\mode{\mathrm {m}}
\def\turb{\mathrm {t}}
\def\up{\mathrm {up}}
\def\dn{\mathrm {dn}}

\def\cm{\mathrm {cm}}
\def\second{\mathrm {s}}
\def\sr{\mathrm {sr}}

\def\eV{\mathrm {eV}}
\def\keV{\mathrm {keV}}
\def\MeV{\mathrm {MeV}}
\def\GeV{\mathrm {GeV}}
\def\Gauss{\mathrm {G}}
\def\milligauss{\mathrm {mG}}
\def\gram{\mathrm {g}}
\def\esu{\mathrm {esu}}
\def\Dlogk{{\mathit \Delta} \! \log_{10} k}

\def\simge{\mathrel{%
   \rlap{\raise 0.511ex \hbox{$>$}}{\lower 0.511ex \hbox{$\sim$}}}}
\def\simle{\mathrel{
   \rlap{\raise 0.511ex \hbox{$<$}}{\lower 0.511ex \hbox{$\sim$}}}}

\newcommand{\expi}[1]{\exp i \left( #1 \right)}

\newcommand{\abs}[1]{\left| #1 \right|}

\slugcomment{}

\shorttitle{Fermi Acceleration in $\log(k)$ space}
\shortauthors{Muranushi et al.}

\begin{document}


\title{Direct Simulations of Particle Acceleration in Fluctuating Electromagnetic Field across a Shock}
%\title{Particle Acceleration at Turbulent Shock, Directly Simulated in Logarithmic Wavenumber Space}


\author{Takayuki Muranushi and Shu-ichiro Inutsuka}
\affil{Department of Physics, Kyoto University,
  Sakyo-ku, Kyoto, 606-8502, Japan; muranushi@tap.scphys.kyoto-u.ac.jp}
%\email{muranushi@tap.scphys.kyoto-u.ac.jp}
%\affil{Theoretical Astrophysics Group, Kyoto University}

%\author{R. J. Hanisch\altaffilmark{5}}
%\affil{Space Telescope Science Institute, Baltimore, MD 21218}


%\altaffiltext{1}{Researching engineer, Preferred Infrastructure.}
%\altaffiltext{1}{Visiting Astronomer, Cerro Tololo Inter-American Observatory.
%CTIO is operated by AURA, Inc.\ under contract to the National Science
%\altaffiltext{2}{Society of Fellows, Harvard University.}
%\altaffiltext{3}{present address: Center for Astrophysics,
%    60 Garden Street, Cambridge, MA 02138}
%\altaffiltext{4}{Visiting Programmer, Space Telescope Science Institute}
%%\altaffiltext{5}{Patron, Alonso's Bar and Grill}


\begin{abstract}

%%Recent advances in understanding of MHD and collisionless turbulence require verification in simulation. 

 We simulate the acceleration processes of collisionless particles in 
a shock structure with magnetohydrodynamical (MHD) fluctuations.
The electromagnetic field is represented as a sum of 
   MHD shock solution ($\Mag_0, \Ele_0$) and 
   torsional Alfven modes spectra ($\delta \Mag, \delta \Ele $).
We represent fluctuation modes in logarithmic wavenumber space.
Since the electromagnetic fields are represented analytically,
our simulations can easily cover as large as eight orders of magnitude 
in resonant frequency,
and do not suffer from spatial limitations of box size or grid spacing.
We deterministically calculate the particle trajectories under the 
Lorenz force for time interval of up to ten years, 
with a time step of $\sim 0.5 \sec$.
This is sufficient to resolve Larmor frequencies without a stochastic treatment.
Simulations show that the efficiency of the first order Fermi acceleration
 can be parametrized by the fluctuation amplitude 
%$m \equiv \log_{10} \eta$
%where
$\eta \equiv \langle \delta B^2 \rangle ^{\frac 1 2 } {B_0}^{-1}$ . 
%We explain physical meaning of this $\eta$ parameter.
Convergence of the numerical results is shown by increasing the number of wave modes in Fourier space 
while fixing $\eta$.

Efficiency of the first order Fermi acceleration has a maximum at
$ \eta \simeq 10^1$.
The acceleration rate depends on the angle between the shock normal and $\Mag_0$,
 and is highest when the angle is zero.
%Angular variation of the acceleration rate has negative correlation with the
%maximum wavelength of the turbulence. 
%This explains observed bipolar structure of SNRs 
%and some shell-like synchrotron-radiating structure.
Our method will provide a convenient tool for comparing collisionless turbulence theories
with, for example, observations of bipolar structure of super nova remnants (SNRs) and shell-like synchrotron-radiating structure.

\end{abstract}


\keywords{acceleration of particles  --- methods: numerical
  --- MHD  --- turbulence}



\section{Introduction} \label{introduction}

Cosmic rays  have the spectrum of
${dN}/{dE} \sim 10^5 ({E}/{\GeV})^{-2.6} 
  { \cm^{-2} \sr^{-1} \second^{-1} \GeV^{-1} }$ up to the so-called
  knee-energy of $10^{15} \eV $.
%(e.g. \citet{2003IJMPA..18.2229A}). 
Cosmic ray propagation theories 
%(e.g. \citet{2005ICRC....3..177K},\citet{2004ApJ...612..238S}) 
suggest ${dN}/{dE} \propto E^{-2}$ energy spectra at the cosmic ray
acceleration sites
\citep[e.g.][]{2007ARNPS..57..285S}.

The current description of cosmic ray acceleration up to knee energy 
$\left(10 ^ {15} \mathrm{eV} \right)$ 
is the well known  first-order Fermi acceleration 
\citep{1977ICRC...11..132A, 1978MNRAS.182..147B}.
In the first-order Fermi acceleration model, 
magnetic turbulence is an important agent for particle acceleration.
Turbulence makes the particle momenta isotropic, 
thus portion of the particles cross the shock front many times.
Expectation value of the kinetic energy after $N_J$ times of shock crossing is
$E\left(N_J\right) = E_0 \left(1+h\right) ^ {N_J}$. 
On the other hand, the probability for a particle to survive $N_J$ shock 
crossing can be roughly estimated as $P\left(N_J\right) = \left(1-p\right) ^ {N_J}$.  
This gives us the power-law spectrum of ${dP}/{dE} = E ^{-(h+p)/h}$.
%Self-consistent generation of turbulence modes for all the required 
%wavelengths, and the isotropization process by the turbulence, remain questionable.

\citet{1996ApJ...473.1029E},
\citet{2000MNRAS.314...65L}, and
\citet{2001MNRAS.321..433B}
 have done simulations to describe the self-consistent generation of turbulence,
with approximations
such as gyro-center approximation, random walk 
approximation, or lowering the dimension. 
On the other hand, the recent development of the particle-in cell simulation has made it
possible to describe the particle acceleration in electron-positron plasma
self-consistently \citep[e.g.][]{2008ApJ...682L...5S}.

In this letter, we propose an alternative approach to the simulation of cosmic
ray acceleration.
We have calculated the motion of particles deterministically, 
solving the particles' cyclotron motion 
from Larmor radii of thermal
particles $\left(\sim 10^9 \cm\right)$ to that of knee energy particles $\left(\sim 10^{17} \cm\right)$.
According to the theories, we assume turbulence spectrum in $\log k$ space.
This allows us to cover a large dynamic-range of space and energy,
which enables us a direct comparison of the accelerated cosmic ray spectra with
the observations.

\section{Numerical Scheme}

\subsection{Representation of Turbulence} \label{representation}

\paragraph{Upstream and Downstream Regions}

In our method, the electromagnetic field and velocity field of a continuous region are given by
\begin{eqnarray}
&\mskip -30mu  \Mag \left(t, \vect r\right) = \Mag_0 + \sum_j {\Mag_{1,j}
  \expi{\vect k_j \cdot \vect r - \omega_j t + \phi_j}} \label{equationB}  \\
&\mskip -30mu  \uel \left(t, \vect r\right)= \uel_0 + \sum_j {\uel_{1,j} \expi{\vect k_j \cdot \vect r - \omega_j t + \phi_j}} \\
&  \Ele \left(t, \vect r\right) = -\frac{1}{c} \uel \left(t, \vect r\right)
\times \Mag \left(t, \vect r\right) \label{equationE}
\end{eqnarray}
where the amplitude and the wavenumber of the $j$-th mode are
\begin{eqnarray}
  \Mag_{ 1,j } &=& (\vect n_1 + i\vect n_2) B_{\turb} \left( \frac {k_j}{k_\max} \right)^{P_{\turb} } \\
  \uel_{ 1,j } &=& \frac{v_{\mathrm A}}{B_0} \Mag_{ 1,j } \\
  k_j&=& {k_\min} \left(\frac {k_\max}{k_\min} \right) ^ {{{j }/\left(
        N_{\mode}-1 \right)}} \label{eq_log_k} \\
  \omega_j &=&  \pm v_{{A}} |{\vect k}_j| + \vect k_j \cdot u
\end{eqnarray}
and the initial phase of the $j$-th mode is  $ \phi_j$.

Here $P_{\turb}$ is the spectral index that reflects  the nature of the turbulence
 and $N_{\mode}$ is the total number of 
the modes ($j \in \{ 1 , \cdots , N_{\mode}\}$),
$\Mag_{1,j}$ is the amplitude for each mode, $\vect k_j$ is its wavenumber,
and $\vect n_1 , \vect n_2$ are two mutually perpendicular unit vectors that
are perpendicular to $k_j$. We choose $\vect k_j$ to be either parallel or
antiparallel to $\Mag_0$.\ADDED  Equation (\ref{eq_log_k})
means that $k_i$ are logarithmically discrete. 

We use $\eta = \left(\Sigma_j {B_{1,j}}^2 \right )^\frac{1}{2}  {B_0}^{-1}$ as the measure of the strength of the fluctuation, independent of $N_{\mode}$.
Because increasing $N_{\mode}$ while  keeping $\eta = const$ 
(1) keeps the magnetic energy in fluctuation mode, and
(2) keeps the expectation value of the fluctuation field 
$ \left | \langle  \Sigma_j {\Mag_{1,j}} \rangle \right |$ the same, if $\phi_j$
are independent. 
We will confirm these properties in section \ref{result}. 

%We have confirmed that changing  $N_{mode}$ while keeping $\eta = const.$ does not affect the results.

The argument to derive $P_{\turb} = -{1}/{3}$ in $\log k$ space is
summarized below: Variables in $\log k$ space are marked by tilde.
The power law energy spectrum is $E(k)dk \propto k^{ -\frac{5}{3}}$ in 
Kolmogorov turbulence case. 
This energy spectrum is in linear bin.
In log energy bin the spectral power is
$ \widetilde E(k)d\log k \equiv kE(k)dk \propto k^{ -\frac{2}{3}}$; and since $\widetilde E = 1/(8 \pi) \widetilde B^2$,
$\widetilde B(k)d\log k \propto {\widetilde E(k)}^{\frac 1 2}d\log k \propto k^{-\frac{1}{3}}$. 
Thus, we get $P_{\turb} = -{1}/{3}$ for our discretization of the turbulent
magnetic field.
%A similar argument leads to $P_{turb} = -\frac{1}{2}$ in
%MHD turbulence \citep{1995ApJ...438..763G} .

\paragraph{Junction Conditions}

We assumed strong shock junction condition with low plasma $\beta$ limit at the shock front:
 $ u_{\dn}            = {J}^{-1} u_{\up}  $,
 $ B_{\parallel  \dn} = B_{\parallel \up}$, and
 $ \Mag_{\perp \dn}   = \Mag_{\perp \up} J $, 
where $J$ is the shock compression ratio, $B _ \parallel$ and
 $\Mag _ \perp$ are components of the $\Mag$ normal and tangential to the
 shock,
respectively.

\begin{figure}
\begin{center}
\includegraphics[scale = 0.50]{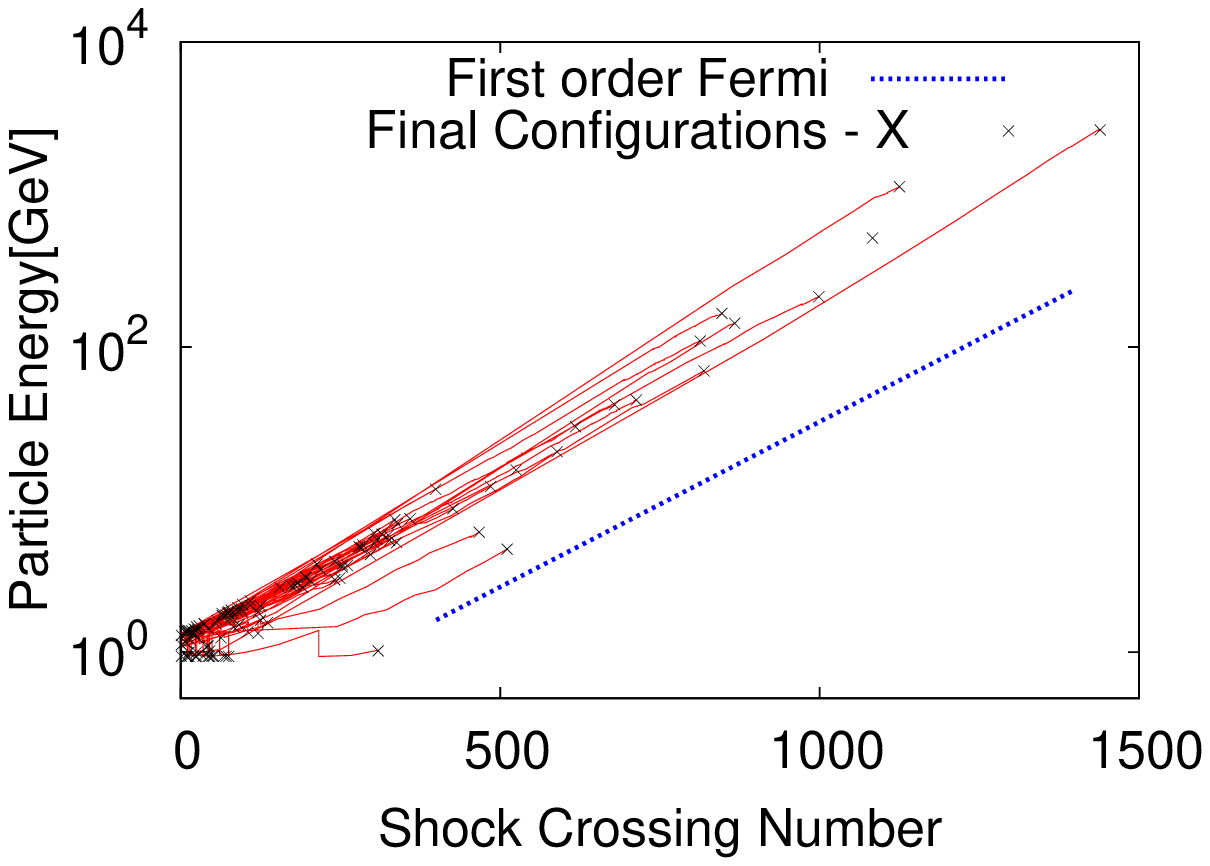}
\caption{
\small
Particle energy as a function of number of shock crossing,
after $\sim 1$ years of time evolution with parameters $\lambda_\max = 10^{17} \cm$ , 
$\eta \equiv  \left(\Sigma_j {B_{1,j}}^2 \right )^\frac{1}{2}  {B_0}^{-1}= 10$, and $\theta = 0$ ($\Mag_0$ is parallel to the shock normal).
The red curves are particle trajectories and 
 inclination of the blue line is the prediction of the first order Fermi acceleration theory.
} \label{jump-energy}

\end{center}
%\end{figure}
%\begin{figure}
%\begin{center}
%\includegraphics[scale=0.480]{energy_for_z_pos_10.eps}
%\caption{
%\small
%Particle energy as function of shock normal particle position.
%Red curves are the traces of particles,
%and black crosses are the final positions and energy of the particles.
%} \label{position-energy}
%\end{center}
%\end{figure}

%\begin{figure}[t]
\begin{center}
\includegraphics[scale=0.56]{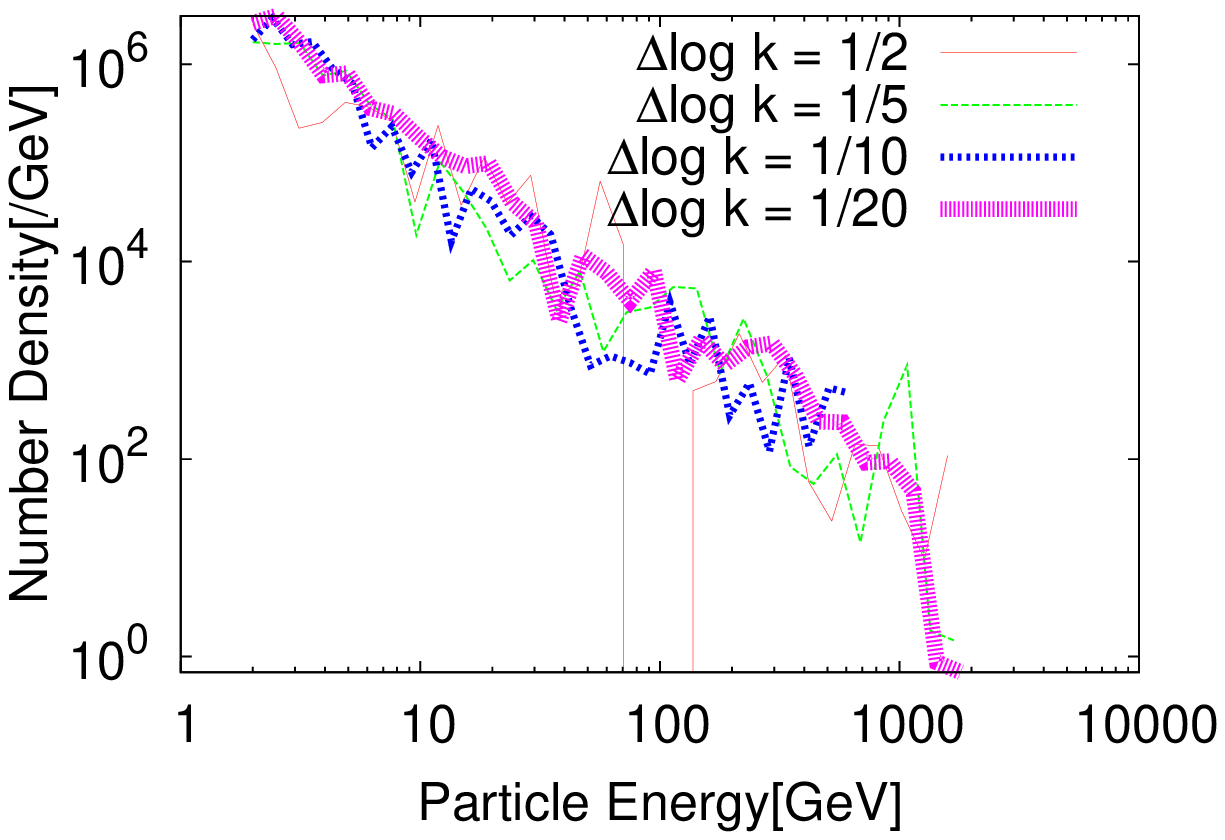}
\caption{
\small 
``Convergence'' test for energy spectrum. 
Curves shows the particle energy spectrum at $\sim 1$ year time evolution. 
Each curve corresponds to discretization of the turbulence spectra into  $\log
k$ space with different
$\Dlogk$ : number of modes per decade,
while $\eta \equiv  \left(\Sigma_j {B_{1,j}}^2 \right )^\frac{1}{2}  {B_0}^{-1}= 10$ is kept.
Other parameters are $\lambda_\max = 10^{17} \cm$ 
 and $\theta = 0$. Particle with energy greater than $2\GeV$ are counted.
}\label{energy-spectrum-convergence}

\end{center}

%\begin{center}
%\includegraphics[scale=0.2]{energy_spectrum.eps}
%\caption{\small
%Energy spectrum for different $\eta$.
%Curves shows the particle energy spectra
%for $\lambda_\max = 10^{17} \cm$ and $\eta = 1$(green), $10$(red), and $300$(blue)
%at $\sim 1$ year time evolution.
%}\label{energy-spectrum}
%\end{center}

%\end{figure}

%\begin{figure}

\begin{center}
\includegraphics[scale=0.56]{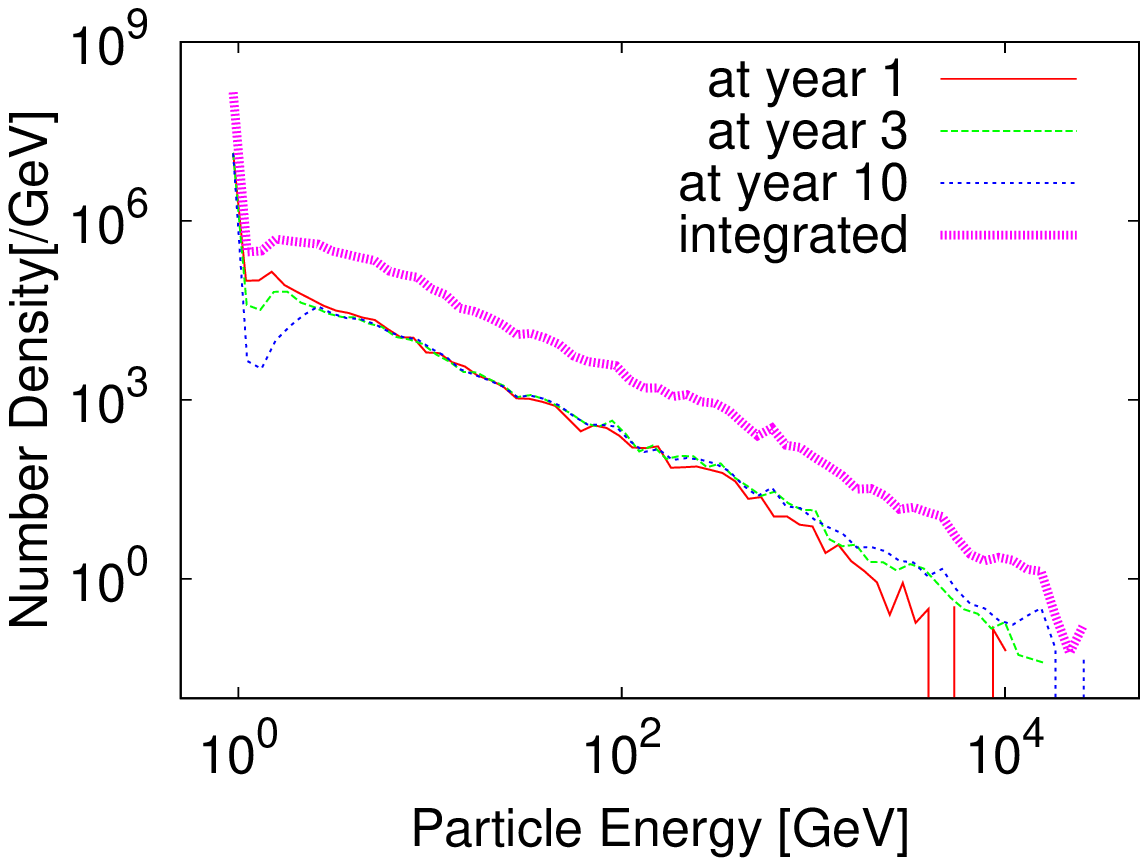}
\caption{\small
Curves show the particle energy spectra
for $\lambda_\max = 10^{17} \cm$ and $\eta = 10$, $\theta = 0$
at $1$, $3$, and $10$ years of time evolution.
Time integrated energy spectrum is shown with the bold curve.
}\label{energy-spectrum-age}

\end{center}

\end{figure}

\subsection{Initial Condition and Equation of Motion}

\paragraph{Initial Condition}

For each set of initial condition we introduce electromagnetic fields
described in section \ref{representation}.
We choose a set of initial turbulence phase $ \{ \phi_j \} $, sign of $\vect
k_j$ and $\omega_j$ from uniform distribution.
Then we put $10^5$ protons in Boltzmann distribution of temperature $T$ at the upstream side of the shock.
\ADDED

We use the values in Table \ref{parameter_table},
based on \citet{2003ApJ...589..827B}'s observation of SN 1006.

\begin{table}[t]
\begin{center}
\caption{Parameters Used for our Simulations}\label{parameter_table}

\footnotesize
\begin{tabular}{cc}
\tableline \tableline
$u_{up} =  3.0 \cdot 10^8 \cm \second^{-1}                 $  & fluid speed in shock frame \\
$v_{A up}=  1.0 \cdot 10^7 \cm \second^{-1}                 $  & Alfven speed in fluid frame \\
$B_{0 up}=  1.0 \cdot 10^{-5} \Gauss                $  & unperturbed magnetic field strength\\
$B_{1 up}= \eta B_{0up}                           $  & torsional Alfven mode energy \\
$\theta                 $  & angle of $\Mag_0$ to shock normal \\
$\lambda_\max = 10^{17} \cm         $  & maximum wavelength of turbulence\\
$\lambda_\min = 10^9 \cm             $  & minimum wavelength of turbulence\\
$T = 0.24 \keV                    $  & temperature of the particles \ADDED \\  
$ m = 1.6 \cdot 10^{-24} \gram$               & particle mass \\
$ e = 4.8 \cdot 10^{-10} \esu$               & particle charge \\
\tableline

\end{tabular} 

\normalsize
\end{center}
\end{table}

%% All your modes are belong to us

\paragraph{Evolution}

We make each turbulence mode propagate at Alfven velocity of the uniform field 
$v_A = {B_0}/{\sqrt{4 \pi \rho}}$ as in Equation \ref{equationB} - \ref{equationE}, 
and updated the particle with 
Lorenz force, with 4-th order Runge-Kutta method. We choose time discretization  $dt$ for each particle at every timestep, so that $dt < 0.1 (1+\eta) e B_0  m^{-1} c^{-1}$ and $dt < 0.03 e \abs{\Mag_0 + \sum \Mag_1}m^{-1} c^{-1}$ always hold. Typical time step is $0.5 \second$ whereas the Larmor period of thermal particle for $B_0$ is $\sim 10^2 \second$. %CHANGE

\begin{eqnarray}
\frac{d \pos}{d t} &=& \frac{\mom}{\gamma m}  \\
\frac{d \mom}{d t} &=& e \left( \Ele + \frac{\vel}{c} \times \Mag \right) 
%\omega_j &=& \vel_A \cdot \vect k_j
\end{eqnarray}

%Since the long wavelength modes of turbulence spectrum contribute as uniform magnetic field,
%the real Alfven velocity $\vec v_A = \frac{B}{\sqrt{4 \pi \rho}}$ will be faster.
%Our underestimation of $v_A$ means smaller $\Ele$ than actual. 
%This may have lead to less acceleration. 
%We have shown that acceleration is still possible with such drawbacks.

%For one background configuration we calculated $\sim 10^5$ particles. 

%We vary the turbulence amplitude
%$\eta = \left( \Sigma_j {B_{1,j}}^2 /  {B_0}^2 \right) ^{\frac 1 2}   $ 
%from $ 1 $ to $300$ ;
%maximum wavelength of turbulence
%$\lambda_\max $ from $ 10^{14} \cm  $ to $ 10^{17} \cm  $  
%and inclination of the unperturbed magnetic field $\theta$ from $0$ to $\frac \pi 2$.

%\be

\section{Result}\label{result}

We have made following examinations for the results of our method.
First, we have traced the particles' energy $ E $ as the function of shock crossing number
 $N_J$ (Figure \ref{jump-energy}).
The inclination of the curves match the inclination\ADDED 
of the theoretical prediction, 
$E(N_J) = \left\{ 1 + (2/3)(v_{up} - v_{dn})/c \right\}^{N_J}$.
Secondly, we have traced the spatial location where the particles gained their kinetic energy. 
We have found that 94\% of the final kinetic energy have been
earned within $1$ final Larmor radius away from the shock. 
This is consistent with the first-order Fermi acceleration picture. 
Thirdly, we have studied the validity of our Fourier representation in $\log k$ space. 
We have kept the physical parameters and increased the number of modes per decade 
$\Dlogk \equiv (N_{\mathrm {mode}} / \log_{10})({k_\max}/{k_\min})$;
we see that the spectra converge, and converge to the theoretical power-law spectrum 
(Figure \ref{energy-spectrum-convergence}). This justifies our use of $ \log
k$ space discretization.

We have done a large number of simulations while varying the background conditions, 
$\lambda_\max$ from $10^{13} \cm$ to $10^{17} \cm$,
$\eta$, the ratio of magnetic energy in fluctuation mode to that in background field from $0.3$ to $300$,
$\theta$, the angle between the background field and the shock normal from $0$ to $\pi/2$.
%Figure \ref{energy-spectrum} shows the energy spectra for different $\eta$. 
%The spectra has the power law dependence $E(k) \propto k^{-2}$
%required by the theory in the range of $1 \leq \eta \leq 10$.
Figure \ref{energy-spectrum-age} shows the time evolution of the energy spectrum for $10$ years.
The high-energy end of the spectrum gradually grows, and reaches $2.5 \times 10^{13} \eV$ by $10$ years.

In our simulation all the particles start its motion in the given time.
Since we don't include the back-reaction from the particles to the
electromagnetic field in our simulations, 
time-integral of energy spectra at each time-slice gives the  steady state energy spectra.
This steady state spectrum is also shown in Figure \ref{energy-spectrum-age}
with thick curve.
The nonthermal component has $E(k) \propto k^{-1.6}$ power-law spectrum that meets the observational requirement
mentioned in section \ref{introduction}.
We can also estimate the ``injection rate'' to be the proportion of particles
that have more than $2 \GeV$ of energy after 1 years.
For $T = 0.24\keV$, $24\keV$, $2.4\MeV$, and $0.24\GeV$, the injection rate was 
${}<0.001$, $1.9 \times 10^{-2}$, $8.4 \times 10^{-2}$, and $0.378$, respectively.\ADDED
The other parameters are $\eta = 30$, $\theta = 0$ and $\lambda_\max = 10^{17} cm$.

%Note, however, that the present model tends to overestimate the injection rate because we use
%infinitesimally thin shock front. In addition, the back reaction of cosmic ray
%particles to the background field modifies the shock structure \citep[e.g.][]{1987MNRAS.225..399F}, thus
%possibly leading to the modification of the energy spectra.
%The rate will be considerably lower in more realistic shock structure where the shock is not a discontinuity but has a scale height.

\begin{figure}[t]
\begin{center}
\includegraphics*[scale=0.48]{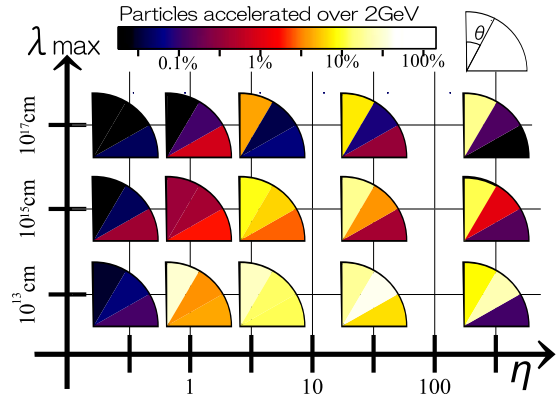}
\caption{\small
Parameter dependence of particle acceleration efficiency .
$\lambda_\max$ is the longest wavelength of the turbulence modes,
$\eta = \left( {\Sigma_j {B_{1,j}}^2 }{ {B_0}}^{-2} \right )^\frac{1}{2} \in \{0.3, 1, 3, 30, 300\}$ is
the ratio of turbulent magnetic field to unperturbed magnetic field,  $\theta$
is the angle between shock normal and $\Mag_0$.  \ADDED
}\label{dsa-map}
\end{center}
\end{figure}

In Figure \ref{dsa-map} we show for all the parameter range the ratio of the
particle numbers that were accelerated to have energy greater than $2
\GeV$. \ADDED
We see that the acceleration is most efficient at polar region ($\theta \simeq
0$) when $\eta > 1$. 
We can understand this dependence of the spectra with background fluid parameters as follows; 
particles are trapped in Larmor motion and tend to move in direction of $\Mag_0$.
Thus particles more easily cross the shockfront when $\Mag_0$ is parallel to shock normal.
If the turbulence amplitude is much weaker, fewer particles get reflected by
pitch angle scattering, and
Fermi acceleration is suppressed.
The injection is more efficient for smaller $\lambda_\max$, 
because more energy is distributed to modes with
smallest wavelengths 
which are resonant with the thermal particles.
\ADDED

% On the other hand, if the amplitude is much stronger, most of the particles get trapped in the turbulence
%and cross the shockfront only once, leading to inefficient acceleration.
% If the maximum turbulence wavelength is large, turbulence modes with largest wavelength (and strongest amplitude) 
% play the role of local $\Mag_0$. Thus we observe almost isotropic Fermi acceleration.

%%%%%%%%%%%%%%%%%%%

If a spherical shock emerges in a uniform mean magnetic field, 
there are two polar region where the mean magnetic field is parallel to the shock normal,
and the equatorial region has the mean magnetic field perpendicular to the shock normal.
Thus, we expect the Fermi acceleration process to be only active in the pair of polar region.
This might explain the bipolar structure we see at SN 1006.

We have also checked the acceleration rate in three-dimensional(isotropic),
 rather than one-dimensional(anisotropic) distribution of ${\vect k}_i$. 
We have found that less significant dependence of injection rate on $\theta$
with larger $\eta$. 
We can interpret this as follows; if the turbulence spectrum is isotropic and the maximum turbulence wavelength is large, turbulence modes with largest wavelength and strongest amplitude 
play the role of local $\Mag_0$. Thus we observe almost isotropic Fermi acceleration.
 This might explain the many SNRs with no typical orientation. \ADDED

\section{Discussion} \label{discussion}

Some might question the validity of $\eta$ value much greater than unity.
However, \citet{2007Natur.449..576U} observed extremely fast varying X-ray images
 at SNR RX J1713.7-3946.
Their observation may indicate that magnetic field is locally enhanced up to $1 {\mathrm
  \milligauss}$ in $\sim 1$ year in SNR, which
corresponds to  $\eta = 100$ case in our model. Our simulations suggest that
such fast-variating spots in SNRs might be the sites of galactic $(E < 10^{15} \mathrm {eV})$ cosmic ray acceleration.

Although we have ignored many of the Fourier modes by adopting $\log k$ space
discretization of the turbulence spectrum\ADDED, 
the validity of the approximation can be argued in many ways.
Most importantly we have confirmed that our measure in $\log k$ space lead to convergence. Figure \ref{energy-spectrum-convergence} shows energy spectra for $\eta = const$, with increasing $\Dlogk$.
Turbulent cascade, from which the very turbulence arises, is by nature a logarithmic process: 
a mode of a certain wavelength couples with the mode of half the wavelength
by nonlinear term of Euler equation.
First order Fermi acceleration also is a logarithmic process: particles gain energy as an exponential function of shock crossing number $E\left(N_J\right) = E_0 \left(1+h\right) ^ {N_J}$.
All these reason combined, waves in logarithmically discretized wavenumber space
act as a sufficient ladder to carry up cosmic ray particles.


%% Included in this acknowledgments section are examples of the
%% AASTeX hypertext markup commands. Use \url without the optional [HREF]
%% argument when you want to print the url directly in the text. Otherwise,
%% use either \url or \anchor, with the HREF as the first argument and the
%% text to be printed in the second.

\acknowledgments

The authors thank S. Inoue and K. Murase for useful comments.
They also thank Center for
Computational Astrophysics (CfCA) of National Astronomical Observatory of
Japan and
Yukawa Institute for Theoretical Physics (YITP) in Kyoto University for
their computing facilities.
The numerical calculations were carried out on Cray XT4 at CfCA and
Altix3700BX2 at YITP.
S. I. is supported
by Grants-in-Aid (15740118, 16077202, and 18540238)
from Ministry of Education, Culture, Sports, Science and Technology
(MEXT) of Japan.
This work was supported by the Grant-in-Aid for the Global COE Program 
"The Next Generation of Physics, Spun from Universality and Emergence" 
from the MEXT of Japan.


%% in parentheses, after the keyword, but they will not be verified.

%%%%%%%{\it Facilities:} \facility{Nickel}, \facility{HST (STIS)}, \facility{CXO (ASIS)}.

%% There should be a \section command for


%%\bibliography{nushio-ApJ.bib}

\begin{thebibliography}{9}
\expandafter\ifx\csname natexlab\endcsname\relax\def\natexlab#1{#1}\fi

\bibitem[{{Axford} {et~al.}(1977){Axford}, {Leer}, \&
  {Skadron}}]{1977ICRC...11..132A}
{Axford}, W., {Leer}, E., \& {Skadron}, G. 1977, in International Cosmic Ray
  Conference, Vol.~11, International Cosmic Ray Conference, 132

\bibitem[{{Bamba} {et~al.}(2003){Bamba}, {Yamazaki}, {Ueno}, \&
  {Koyama}}]{2003ApJ...589..827B}
{Bamba}, A., {Yamazaki}, R., {Ueno}, M., \& {Koyama}, K. 2003, \apj, 589, 827

\bibitem[{{Bell}(1978)}]{1978MNRAS.182..147B}
{Bell}, A. 1978, \mnras, 182, 147

\bibitem[{{Bell} \& {Lucek}(2001)}]{2001MNRAS.321..433B}
{Bell}, A., \& {Lucek}, S. 2001, MNRAS, 321, 433

\bibitem[{{Ellison} {et~al.}(1996){Ellison}, {Baring}, \&
  {Jones}}]{1996ApJ...473.1029E}
{Ellison}, D., {Baring}, M., \& {Jones}, F. 1996, \apj, 473, 1029

\bibitem[{{Lucek} \& {Bell}(2000)}]{2000MNRAS.314...65L}
{Lucek}, S., \& {Bell}, A. 2000, \mnras, 314, 65

\bibitem[{{Spitkovsky}(2008)}]{2008ApJ...682L...5S}
{Spitkovsky}, A. 2008, \apjl, 682, L5

\bibitem[{{Strong} {et~al.}(2007){Strong}, {Moskalenko}, \&
  {Ptuskin}}]{2007ARNPS..57..285S}
{Strong}, A.~W., {Moskalenko}, I.~V., \& {Ptuskin}, V.~S. 2007, Annual Review
  of Nuclear and Particle Science, 57, 285

\bibitem[{{Uchiyama} {et~al.}(2007){Uchiyama}, {Aharonian}, {Tanaka},
  {Takahashi}, \& {Maeda}}]{2007Natur.449..576U}
{Uchiyama}, Y., {Aharonian}, F., {Tanaka}, T., {Takahashi}, T., \& {Maeda}, Y.
  2007, \nat, 449, 576

\end{thebibliography}

\end{document}